\documentclass[aps,prl,superscriptaddress,amsmath,amssymb,twocolumn,floatfix,nofootinbib]{revtex4-1}
\usepackage{xcolor}
\usepackage{graphicx}
\usepackage[colorlinks=true,citecolor=blue,linkcolor=blue,breaklinks=true]{hyperref}
\usepackage{natbib}

\begin{document}

\title{Searching for Afterglow: Light Dark Matter Boosted by Supernova Neutrinos}

\affiliation{Institute of Physics, Academia Sinica, Taipei 115, Taiwan}
\affiliation{Department of Physics, National Taiwan University, Taipei 106, Taiwan}
\affiliation{Institute of Astronomy and Astrophysics, Academia Sinica, Taipei 106, Taiwan}
\affiliation{Physics Division, National Center for Theoretical Sciences, Taipei 106, Taiwan}

\author{Yen-Hsun Lin}
\email{yenhsun@phys.ncku.edu.tw}
\affiliation{Physics Division, National Center for Theoretical Sciences, Taipei 106, Taiwan}
\affiliation{Institute of Physics, Academia Sinica, Taipei 115, Taiwan}

\author{Wen-Hua Wu}
\affiliation{Department of Physics, National Taiwan University, Taipei 106, Taiwan}
\affiliation{Institute of Physics, Academia Sinica, Taipei 115, Taiwan}

\author{Meng-Ru Wu}
\email{mwu@gate.sinica.edu.tw}
\affiliation{Institute of Physics, Academia Sinica, Taipei 115, Taiwan}
\affiliation{Institute of Astronomy and Astrophysics, Academia Sinica, Taipei 106, Taiwan}
\affiliation{Physics Division, National Center for Theoretical Sciences, Taipei 106, Taiwan}

\author{Henry Tsz-King Wong}
\email{htwong@phys.sinica.edu.tw}
\affiliation{Institute of Physics, Academia Sinica, Taipei 115, Taiwan}

\begin{abstract}

A novel analysis is performed, incorporating \emph{time-of-flight} (TOF) information to study the
interactions of dark matter (DM) with standard model particles. After supernova
(SN) explosions, DM with mass $m_\chi\lesssim\mathcal{O}({\rm MeV})$ in the halo can be boosted by SN 
neutrinos (SN$\nu$) to relativistic speed. The SN$\nu$ boosted DM (BDM) arrives on Earth with TOF
which depends only on $m_\chi$ and is independent of the cross section. These BDMs can
interact with detector targets in low-background experiments and manifest as
\emph{afterglow} events after the arrival of SN$\nu$. The characteristic TOF spectra of the BDM
events can lead to large background suppression and unique determination of $m_\chi$.
New cross section constraints on $\sqrt{\sigma_{\chi e} \sigma_{\chi\nu}}$ are derived from SN1987a in the Large
Magellanic Cloud with data from the Kamiokande and Super-Kamiokande
experiments. Potential sensitivities for the next galactic SN with Hyper-Kamiokande
are projected.
This analysis extends the existing bounds on $\sqrt{\sigma_{\chi e}\sigma_{\chi \nu}}$ over a broad range of $r_\chi=\sigma_{\chi \nu}/\sigma_{\chi e}$. In particular, the improvement is by 1--3 orders of magnitude for $m_\chi<\mathcal{O}(100\,{\rm keV})$ for $\sigma_{\chi e}\sim\sigma_{\chi \nu}$.
Prospects of exploiting TOF information in other astrophysical
systems to probe exotic physics with other DM candidates are discussed.

\end{abstract}
\maketitle

{\it Introduction.---}Although there is compelling evidence on the existence of dark matter
(DM) as an additional gravity source, its properties and interactions remain unknown~\cite{Battaglieri:2017aum,Workman:2022}. Experimental searches of DM are intensely pursued worldwide~\cite{AMS:2015azc,LUX:2016ggv,Fermi-LAT:2017opo,LUX:2017ree,XENON:2018voc,DAMPE:2017fbg,XENON:2019gfn,XENON:2019zpr,SENSEI:2019ibb,SuperCDMS:2018mne}. Direct
detection (DD) experiments focus on the weakly interacting massive particle (WIMP)
scenario of DM mass $m_\chi \gtrsim \mathcal{O}({\rm GeV})$, with sensitivities approaching the neutrino floor~\cite{Workman:2022}. The search for lighter WIMPs is an active area of research. One scenario with
rapidly expanding interest is where light DM is upscattered or boosted by known
cosmic particles including baryons, electrons, and neutrinos~\cite{Bringmann:2018cvk,Ema:2018bih,Cappiello:2019qsw,Dent:2019krz,Wang:2019jtk,Zhang:2020nis,Guo:2020drq,Ge:2020yuf,Cao:2020bwd,Jho:2020sku,Cho:2020mnc,Lei:2020mii,Guo:2020oum,Xia:2020apm,Dent:2020syp,Ema:2020ulo,Flambaum:2020xxo,Jho:2021rmn,Das:2021lcr,Bell:2021xff,Chao:2021orr,Ghosh:2021vkt,Feng:2021hyz,Wang:2021nbf,Xia:2021vbz,Wang:2021jic,PandaX-II:2021kai,CDEX:2022fig,Granelli:2022ysi}. The boosted
DM (BDM) then carries kinetic energy $T_\chi$ much larger than when it is nonrelativistic
(with velocity $v_\chi \sim 10^{-3}$) according to the Halo model. Nuclear and electron recoil
events from BDM interaction with the detector targets will therefore have increased
energy deposition, making DM with $m_\chi \lesssim \mathcal{O}({\rm GeV})$ experimentally accessible.

\emph{Time-of-flight} (TOF) techniques are matured laboratory tools for differentiation or
measurement of particle masses. This technique, however, has not been well
exploited to probe exotic physics in astrophysical systems. One notable exception is
the neutrino mass constraints derived from the timing distributions of supernova neutrinos
(SN$\nu$s) from SN1987a~\cite{Arnett1987,Kolb1987}. We explore in this Letter a novel scenario of
BDM with kinetic energy injected by SN$\nu$ interactions, and in particular where the
prompt SN$\nu$ burst is also detected, providing a time-zero definition in terrestrial
experiments. The prompt SN$\nu$ events will be followed by time-evolving BDM
afterglow events where energy and time can be measured. The delay time between
BDM and SN$\nu$ is a distinctive ``smoking-gun'' signature and provides unique
information to infer $m_\chi$, independent of the interaction cross section. Specifically, a
delay time of $\Delta t\simeq 10$~days$\times [R/(8~{\rm kpc})][m_\chi/(10~{\rm keV})]^2[T_\chi/(10~{\rm MeV})]^{-2}$ for SN$\nu$BDM
traveling an astronomical distance $R$ before reaching the Earth highlights that
although BDM has $v_\chi\sim c$, the delay can be substantial but measurable in a duration
post the arrival of SN$\nu$. In contrast, most proposed BDM scenarios rely on steady
sources, e.g., cosmic rays \cite{Bringmann:2018cvk,Ema:2018bih,Cappiello:2019qsw,Dent:2019krz,Wang:2019jtk,Guo:2020drq,Ge:2020yuf,Cao:2020bwd,Jho:2020sku,Cho:2020mnc,Lei:2020mii,Guo:2020oum,Xia:2020apm,Dent:2020syp,Ema:2020ulo,Flambaum:2020xxo,Bell:2021xff,Feng:2021hyz,Wang:2021nbf,Xia:2021vbz}, stellar $\nu$~\cite{Zhang:2020nis,Jho:2021rmn}, diffuse SN$\nu$~\cite{Das:2021lcr,Ghosh:2021vkt}, etc, for which the BDM flux is constant with time and lacks any time-dependent feature.

We explore the signatures of SN$\nu$BDM with SN1987a in the Large
Magellanic Cloud~(LMC) and a future supernova (SN) in the Galactic
Center~(GC) to derive the fluxes and the associated electron-recoil event rates via $\sigma_{\chi e}$ in multikiloton water Cherenkov detectors,
including Kamiokande, Super-Kamiokande (Super-K), and Hyper-Kamiokande. 
The scenario of SN$\nu$BDM depends on finite DM cross section with $\nu$ ($\sigma_{\chi\nu}$), which
may originate from an effective Lagrangian $\bar{\chi}\Gamma{\chi}\bar{\ell_i}\Gamma\ell_i/\Lambda$ where $\chi$ and $\ell_i=(\nu_i,i)$ are the DM and SM fields with $i=e,\mu,\tau$. The vertex $\Gamma$ denotes the interaction type and $\Lambda$ indicates certain cutoff scale.
Possible interactions between $\chi$ with $\nu$ is a subject of intense recent interest~\cite{Holdom:1985ag,Arguelles:2017atb,Chang:2018rso,Murase:2019xqi,Lin:2021hen,Croon:2020lrf,Escudero:2019gzq,Foldenauer:2018zrz}.
They can naturally arise in many particle physics models such as
the extensively studied $B-L$ and $L_\mu-L_\tau$,
where the new gauge bosons can kinematically mix with the standard model photon.
Further constraints will be provided by this work.

\begin{figure}
\begin{centering}
\includegraphics[width=0.8\columnwidth]{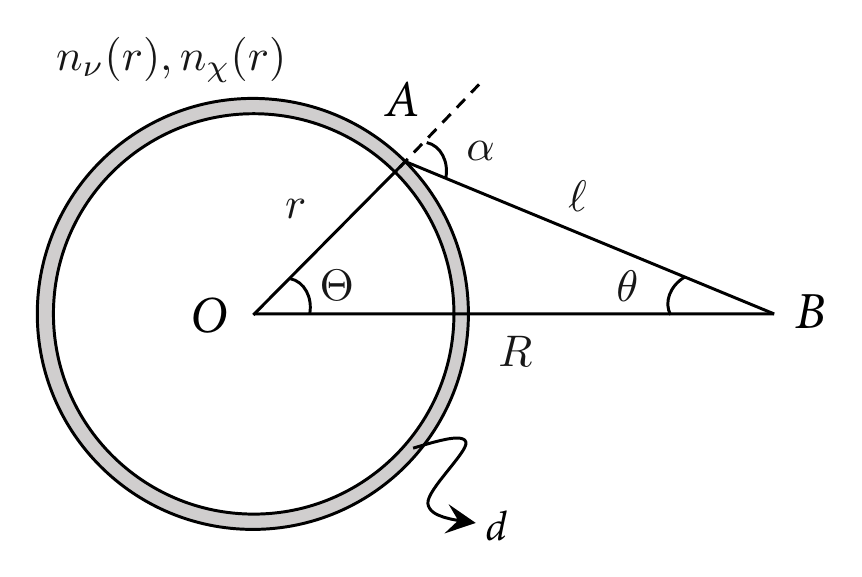}
\end{centering}
\caption{\label{fig:scheme}
Schematic plot of DM boosted by SN$\nu$ within an expanding spherical shell with width $d$ at radius $r$.
The SN occurs at $O$.
BDM from $A$ arrives $B$ with an scattering angle $\alpha$.
}
\end{figure}

{\it DM boosted by SN$\nu$.---}Assuming a SN explodes near the center of a galaxy~(location $O$ 
in Fig.~\ref{fig:scheme}), it emits a large amount of $\mathcal{O}(10)$~MeV neutrinos within $\tau \approx 10$~s carrying total luminosity $L_{\nu,{\rm tot}}\approx 3\times 10^{52}$~erg~s$^{-1}$. 
We approximate these SN$\nu$ by an expanding thin spherical shell with a radius $r$ away from $O$ and a thickness $d\approx c\tau$ 
(see Fig.~\ref{fig:scheme}). 
The radially propagating SN$\nu$ within the shell has a number density of 
\begin{equation}\label{eq:dnnude}
\frac{dn_{\nu}}{dE_{\nu}} =\sum_{i}\frac{L_{\nu_{i}}}{4\pi r^{2}\langle E_{\nu_{i}}\rangle}E_{\nu}^{2}f_{\nu_{i}}(E_{\nu}), 
\end{equation}
where $L_{\nu_{i}}=L_{\nu,{\rm tot}}/6$ 
is the luminosity 
of each flavor ($\nu_e$,~$\nu_\mu$,~$\nu_\tau$ and their antineutrinos).  
We take the average energy $\langle E_{\nu_e}\rangle$,~$\langle E_{\bar\nu_e}\rangle$, 
and~$\langle E_{\nu_x}\rangle$ ($\nu_x\in \{ \nu_\mu, \nu_\tau, \bar\nu_\mu, \bar\nu_\tau\}$) to be 11,~16,~25~MeV, respectively~\cite{Duan:2006an}. 
The energy distribution follows a Fermi-Dirac distribution $f_{\nu_{i}}$ with a pinch parameter $\eta_{\nu_i}\equiv\mu_{\nu_i}/T_{\nu_i}=3$, such that $T_{\nu_i} \approx \langle E_{\nu_i}\rangle/3.99$.

\begin{figure}
\begin{centering}
\includegraphics[width=1\columnwidth]{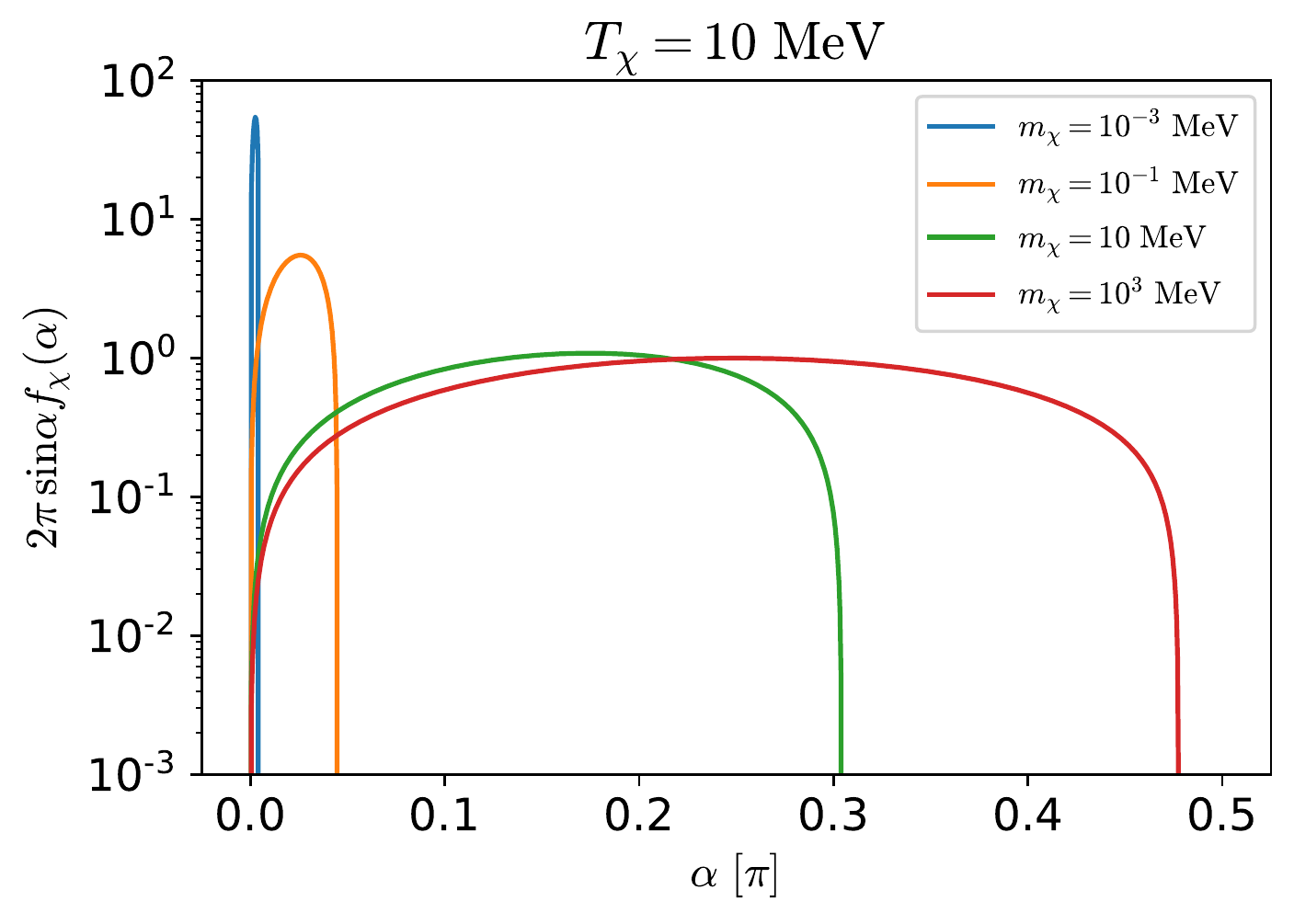}
\end{centering}
\caption{\label{fig:f}
The BDM angular distribution 
$f_\chi(\alpha)$ times $2\pi\sin\alpha$ for 
fixed $T_\chi=10$~MeV 
and different $m_\chi$.}
\end{figure}

With a nonvanishing DM-$\nu$ interaction, these neutrinos can upscatter DM in the halo [with number density $n_\chi(r)$] when they propagate outward. 
The BDM from location $A$ can reach the Earth at $B$ (with a distance $R$ away from the center) with a scattering angle $\alpha$ after traveling a length $\ell$. 
At neutrino energy
$E_\nu$ much larger than the typical DM kinetic energy in the halo,
DM can be approximated as at rest, and the BDM kinetic energy is given by
\begin{equation}
T_{\chi}=\frac{E_{\nu}^2}{E_{\nu}+m_{\chi}/2}\left(\frac{1+\cos\theta_{c}}{2}\right)\label{eq:Tx},
\end{equation}
where $\theta_{c}\in[0,\pi]$ is the scattering angle in the center-of-mass (c.m.) frame.
One can relate $\theta_{c}$ to the lab frame scattering angle $\alpha\in[0,\pi/2]$
by $\theta_{c}=2\tan^{-1}(\gamma\tan\alpha)$
and $\gamma=(E_{\nu}+m_{\chi})/\sqrt{m_{\chi}(2E_{\nu}+m_{\chi})}$.
Assuming $\sigma_{\chi\nu}$ is independent of $\theta_c$ in 
the c.m.~frame, the normalized BDM angular distribution in the lab frame is given by 
\begin{equation}
f_\chi(\alpha,E_{\nu})= \frac{\gamma^2\sec^{3}\alpha}{\pi(1+\gamma^{2}\tan^{2}\alpha)^{2}},\label{eq:f}
\end{equation}
such that $\int d\Omega_\alpha f_\chi (\alpha,E_\nu)=1$ for any given $E_\nu$, where $d\Omega_\alpha=2\pi \sin\alpha d\alpha$.
In Fig.~\ref{fig:f}, we plot $2\pi\sin\alpha f_\chi (\alpha)$ for a fixed
$T_{\chi}=10\,{\rm MeV}$ (corresponding to different $E_\nu$) with different $m_{\chi}$. 
It shows that for BDM with $m_\chi/T_\chi\ll 1$, they are 
confined within a small scattering angle relative to the direction of SN$\nu$.

The BDM emissivity $j_{\chi}$ at location $A$ can be written as
\begin{equation}
j_{\chi}(r,T_{\chi},\alpha)=c\sigma_{\chi\nu}n_{\chi}
\left(\frac{dn_{\nu}}{dE_{\nu}}\right)
\left(\frac{dE_{\nu}}{dT_{\chi}}\frac{v_\chi}{c}\right)f_{\chi},
\label{eq:emissivity}
\end{equation}
where the BDM velocity $v_{\chi}/c=\sqrt{T_{\chi}(2m_{\chi}+T_{\chi})}/(m_{\chi}+T_{\chi})$, and can be evaluated using Eqs.~\eqref{eq:dnnude} to \eqref{eq:f}.

{\it Time-dependent BDM flux at Earth.---}To obtain the BDM flux (number of BDM per unit time per unit energy per solid angle) at Earth 
$d\Phi_\chi/(dT_\chi d\Omega)$ 
(location $B$ in Fig.~\ref{fig:scheme}),
we shall integrate all 
$j_{\chi}$ along the line of sight $\ell$, 
\begin{align}
\frac{d\Phi_{\chi}(T_\chi,\theta,t^\prime)}{dT_{\chi}d\Omega} &=\int d\ell j_{\chi}(r,T_{\chi},\alpha)
H\left(t^{\prime}-\frac{r}{c}-\frac{\ell}{v_{\chi}}\right)\nonumber \\
  & \quad \times H\left(\frac{r}{c}+\frac{\ell}{v_{\chi}}+\tau-t^{\prime}\right)\label{eq:los_int},
\end{align}
where $d\Omega=2\pi\sin\theta d\theta$ is viewed from $B$.
The Heaviside functions limit 
$j_\chi$ to being nonzero only within the spherical shell of width $d$ where SN$\nu$ are present.
The arrival time of BDM, $t^{\prime}$, relative to the time of SN explosion, includes the propagation time of SN$\nu$ from $O$ to $A$ ($r/c$) and 
the traveling time of BDM from $A$ to $B$ ($\ell/v_{\chi}$).

Integrating Eq.~\eqref{eq:los_int} over $d\Omega$ and approximating $ H(x-x_{0})H(x_{0}+\epsilon-x)\sim \epsilon \delta(x_0)$ for
$\epsilon \ll x_{0}$, we obtain
\begin{align}
\frac{d\Phi_{\chi}(T_\chi, t^\prime)}{dT_{\chi}} &=2\pi\tau\int d\cos\theta d\ell  j_{\chi}(r,T_{\chi},\alpha)\delta\left(t^{\prime}-\frac{r}{c}-\frac{\ell}{v_{\chi}}\right) 
\nonumber \\
&=\left.2\pi\tau\int_{0}^{1}d\cos\theta \mathcal{J} j_{\chi}(r,T_{\chi},\alpha)\right|_{t^{\prime}=\frac{r}{c}+\frac{\ell}{v_{\chi}}},\label{eq:BDM_flux}
\end{align}
where
\begin{equation}
    \mathcal{J}=\left(\frac{\ell-R\cos\theta}{rc}+\frac{1}{v_{\chi}}\right)^{-1}\label{eq:J}
\end{equation}
appears due to the change of variable $d\ell = \mathcal{J}dt^\prime$.
Note that for a given $(t^\prime,\theta)$, one can find a
unique solution of $(r,\ell,\alpha)$ and compute the integration.

\begin{figure}
\begin{centering}
\includegraphics[width=1\columnwidth]{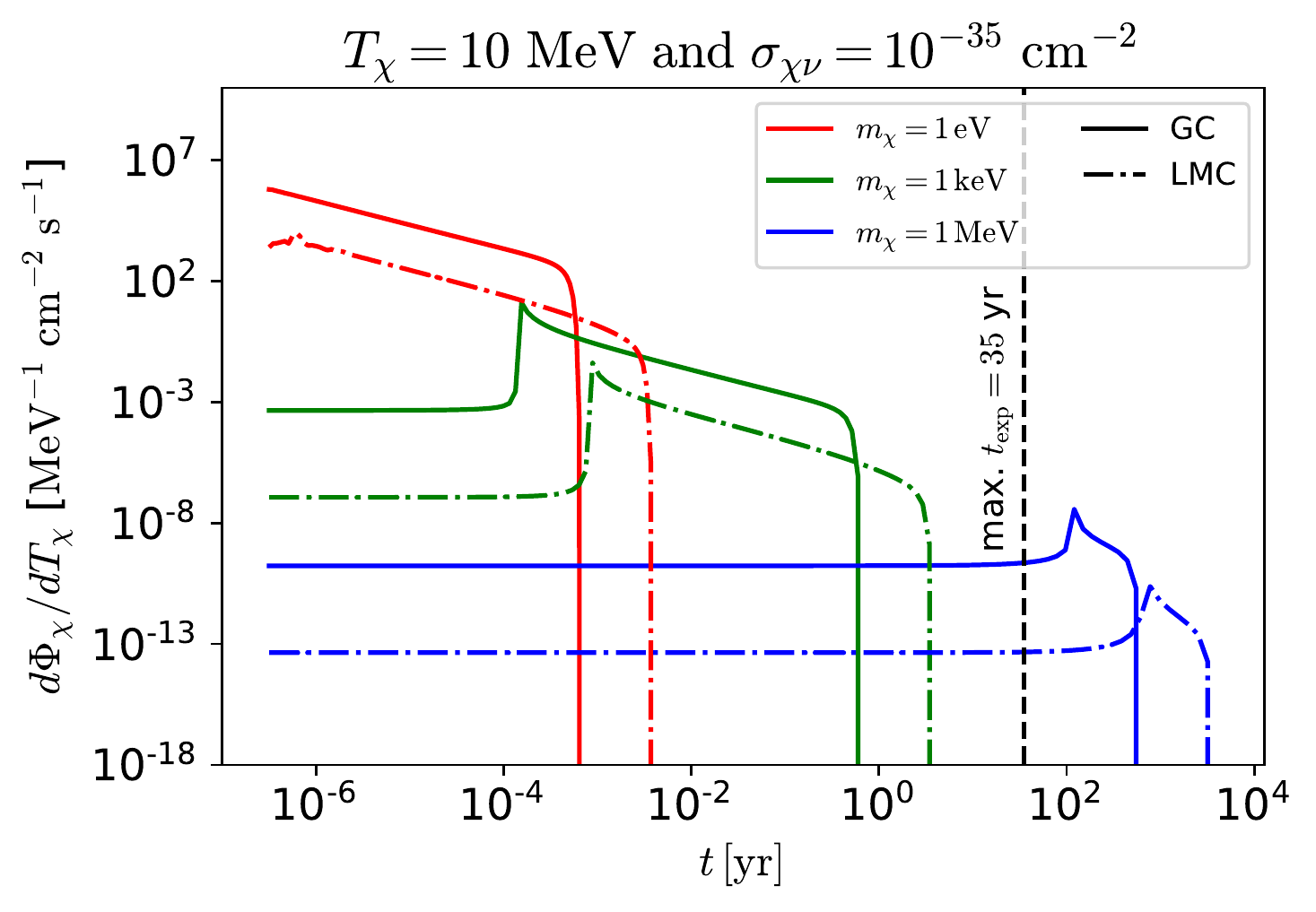}
\par\end{centering}
\caption{\label{fig:diff_BDM_flux}The BDM flux at Earth vs.~$t$ with
different $m_{\chi}$ for 
$T_{\chi}=10\,{\rm MeV}$ and $\sigma_{\chi\nu}=10^{-35}\,{\rm cm^2}$. 
Fluxes resulting from a SN in the GC and from SN1987a in LMC are shown with solid and dash-dotted lines.
The black dashed line indicates the maximum exposure time $t_{\rm exp}=35$~years (see text for details).
}
\end{figure}

{\it BDM flux from SN in the GC and LMC.---}We now compute the BDM fluxes at the Earth from SN1987a in LMC and from a SN in the GC. 
We characterize $n_\chi$ in the Milky Way (MW) and LMC by Navarro-Frenk-White~(NFW) and Hernquist profiles respectively. 
Both share the same expression
\begin{equation}
n_{\chi}(r)=\frac{\rho_{s}}{m_{\chi}}\frac{1}{\frac{r}{r_{s}}(1+\frac{r}{r_{s}})^{n}},\label{eq:nx}
\end{equation}
with $(n,\rho_{s},r_{s})=(2,184\,{\rm MeV\,cm^{-3}},24.4\,{\rm kpc})$ for MW~\cite{Bertone:2004pz} and
$(n,\rho_{s},r_{s})=(3,68\,{\rm MeV\,cm^{-3}},31.9\,{\rm kpc})$ for LMC~\cite{Erkal_2019}.
The distances $R$ for the two are $(R_{\rm GC},R_{\rm LMC})=(8.5,50)\,{\rm kpc}$.
We neglect the contribution from $r<10^{-5}\,{\rm kpc}$ since the profile in the inner region is highly uncertain and the adopted profile diverges when $r\rightarrow 0$.

Fig.~\ref{fig:diff_BDM_flux} shows $d\Phi_{\chi}/dT_{\chi}$ versus $t>\tau$ for $T_\chi=10$~MeV with different $m_{\chi}$ for SN in the GC (solid) and in LMC (dot-dashed), assuming $\sigma_{\chi\nu}=10^{-35}$~cm$^2$.
Note that we define a shifted time coordinate
$t = t^\prime - R/c $ as the delayed arrival time for BDM relative to SN$\nu$.
For $m_\chi=1$~keV and $1$~MeV, 
the most prominent feature is that the BDM fluxes contain a rising part and peak at $t_p\approx R(1/v_{\chi}-1/c)$.
This is mainly due to the increase of $n_\chi\propto r^{-1}$ toward the halo center. 
The postpeak tails 
are due to BDM contributions with larger scattering angles.
For $m_\chi=1\,{\rm eV}$, $t_p\approx 0.004\,{\rm s}$ is too short and overlaps with the 10~s duration of SN$\nu$ to be shown in Fig.~\ref{fig:diff_BDM_flux}. 
Comparing BDM fluxes coming from the GC to LMC, the LMC cases have smaller fluxes and larger $t_p$ 
due to larger $R$ and smaller halo density.

Fig.~\ref{fig:diff_BDM_flux} also shows another important feature---the BDM flux for a given $T_\chi$ and $m_\chi$ vanishes after some time post $t_p$, which is related to the sharp cutoff of $f_\chi$ shown in Fig.~\ref{fig:f}.
This allows us to consider a reduced duration for BDM searches after the arrival of SN$\nu$.
Practically, a detector that can probe BDM has a threshold energy $T_{\rm th}$, below which the detector is insensitive to BDM. 
Thus, for a given $m_\chi$, one can define the latest possible arrival time of BDM with $T_\chi=T_{\rm th}$ as the vanishing time $t_{\rm van}$ to analyze the data.
We stress that all these time-dependent features only depend on $m_\chi$ but not $\sigma_{\chi\nu}$. 
Consequently, if such BDM is detected,
analyzing the time profile of the
signal will allow direct measurements of $m_\chi$.

{\it Events in Kamiokande and Super-K.---}For BDM that also interact with electrons with a cross section $\sigma_{\chi e}$, they can produce signals in neutrino or DM experiments. 
The total event number $N_\chi$ induced by BDM with $T_{\rm th}\leq T_\chi\leq T_{\rm max}$ within an exposure time $t_0\leq t\leq t_{\rm exp}$ is given by
\begin{equation}
N_{\chi}=N_{e}\sigma_{\chi e}\int^{T_{\rm max}}_{ T_{\rm th}}dT_{\chi}\int_{t_0}^{t_{\rm exp}} dt~ \epsilon \frac{d\Phi_{\chi}}{dT_{\chi}},\label{eq:Nx}
\end{equation}
with $N_{e}$ 
the total target number of electrons and $\epsilon$ the signal efficiency.
We consider
the water Cherenkov experiments, Kamiokande and Super-K, to calculate $N_\chi$
for BDM from LMC (by SN1987a) and from a SN in GC. 
They have $N_e=(M_T/m_{\rm H_2 O})N_A n_e$ with $M_T$ the fiducial detector mass, 
$m_{\rm H_2 O}$ the water molar mass, $N_A$ the Avogadro constant and $n_e$ the electron number per water molecule.
We take $M_T=2.2$ and $22.5$ kton for Kamiokande and Super-K, respectively~\cite{Kamiokande1987,Super-Kamiokande:2016yck}, and set $(T_{\rm th},T_{\rm max})=(5,100)\,{\rm MeV}$ for both.
We make a conservative choice of taking $\epsilon=50\%$, lower than the energy-dependent efficiency roughly ranging from 50\% to 75\% reported for solar $\nu$ detection in Super-K \cite{Super-Kamiokande:2016yck}.
For signal duration, we consider $t_0=10\,{\rm s}$ to approximately exclude 
events produced by SN$\nu$, and let $t_{\rm exp}={\rm min}(t_{\rm van}, t_{\rm cut}=35~\rm{yrs})$ depending on $m_\chi$.
For the LMC case, the considered duration thus includes the running time of Kamiokande from
1987 to 1996 and Super-K after 1996 for heavier $m_\chi$. 
For the GC case, we consider Super-K only.
The main background for both comes from the solar 
and atmospheric neutrinos for $T_\chi\lesssim 20\,{\rm MeV}$ and $T_\chi\gtrsim 20\,{\rm MeV}$, respectively.
We adopt values in Table XIV of Ref.~\cite{Super-Kamiokande:2016yck} and the FLUKA simulation result in Ref.~\cite{Battistoni:2005pd} to estimate the background.

\begin{figure}
\begin{centering}
\includegraphics[width=1\columnwidth]{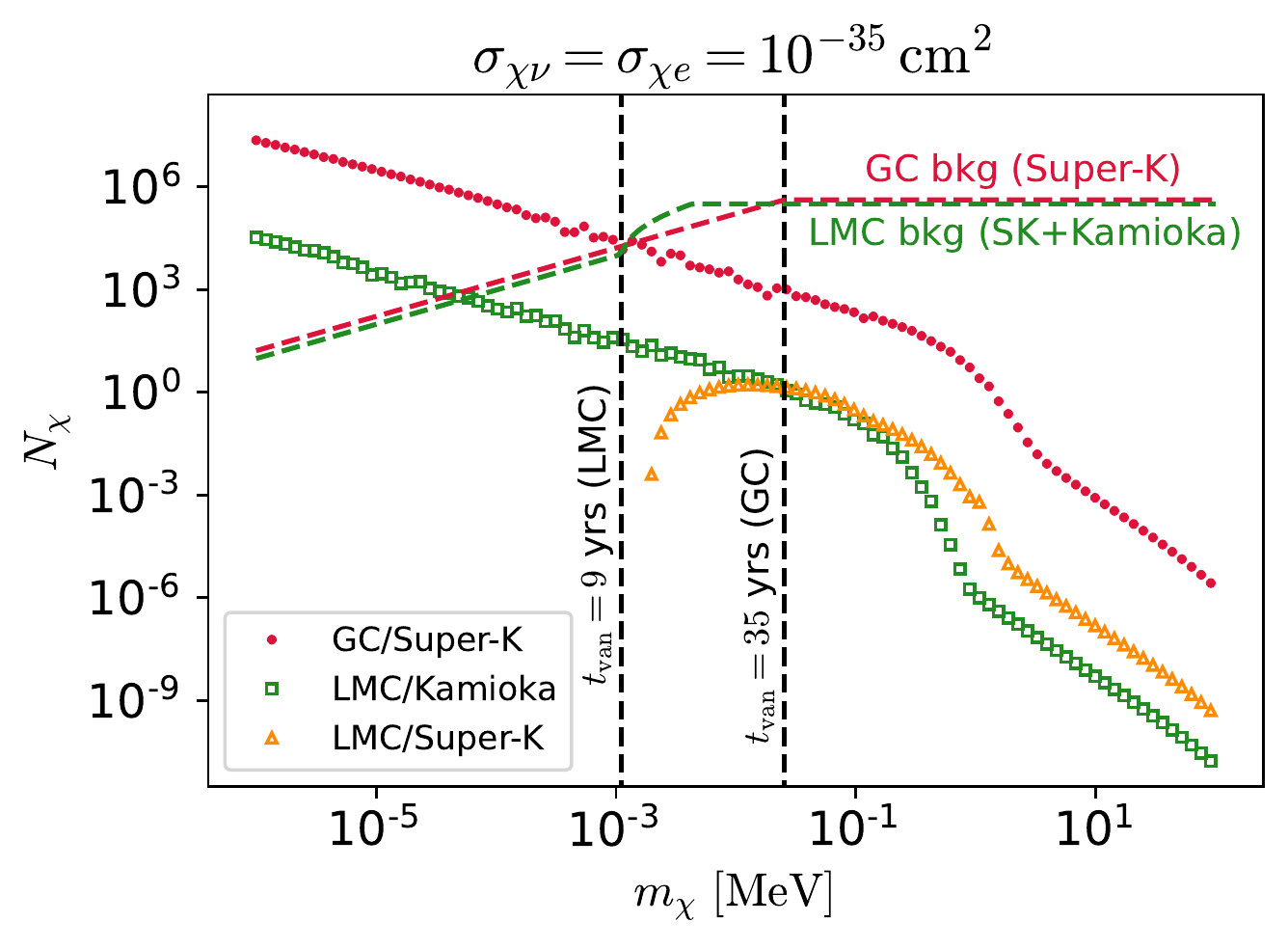}
\par\end{centering}
\caption{\label{fig:int_BDM_event}BDM events in water Cherenkov detectors $N_\chi$ as a function of $m_\chi$ for both the GC and LMC (SN1987a) cases. 
For LMC, events in Kamiokande (hollow-green squares) and Super-K (hollow-orange triangles) are shown separately. 
For the GC, only Super-K (red dots) is considered.
Background counts (dashed lines) are also shown for both cases.
}
\end{figure}

Fig.~\ref{fig:int_BDM_event} shows $N_\chi$ vs.~$m_\chi$ resulting from the GC and LMC given $\sigma_{\chi\nu}=\sigma_{\chi e}=10^{-35}$~cm$^2$.
We first discuss the GC case where only Super-K is considered.
The red-solid dots show that $N_\chi\propto m_\chi^{-1}$ perfectly for $m_\chi\leq 25$~keV, which corresponds to having $t_{\rm exp}=t_{\rm van}\leq t_{\rm cut}$.
This is because for smaller $m_\chi$, all BDM arrive at the 
detector before $t_{\rm cut}$ so that $N_\chi$ is proportional to the amount of DM in the halo.
For heavier $m_\chi$, however, a larger part of BDM flux only arrives after $t_{\rm cut}$ (see Fig.~\ref{fig:diff_BDM_flux}), leading to a faster decrease of $N_\chi$ with increasing $m_\chi$.
For the same reason, the background counts (red-dashed curve) $N_b\simeq 526 M_T t_{\rm exp}$ for $m_\chi\leq 25$~keV due to a constant background rate of $\sim 526$~events per kton per year~\cite{Super-Kamiokande:2016yck}. 
For $m_\chi > 25$~keV where $t_{\rm exp}=t_{\rm cut}$ is applied, $N_b$ stays constant.

For BDM associated with SN1987a in LMC, we plot $N_\chi$ in Kamiokande (1987--1996) and Super-K (after 1996) by hollow green squares and orange triangles separately. 
The behavior of $N_\chi(m_\chi)$ in Kamiokande is similar to that of the GC case, but falls off faster for large $m_\chi$ due to the maximal exposure time of 9 yr only. 
The difference at small $m_\chi$ is mainly due to different detector fiducial mass $M_T$, geometric dilution factor $1/R^{2}$, and the characteristic density $\rho_s$ of DM profiles. 
A simple estimate gives $N_\chi^{\rm GC}/N_\chi^{\rm LMC}\sim(\rho_s^{\rm NFW}/\rho_s^{\rm Hernquist})(R_{\rm LMC}^2/R_{\rm GC}^2) (M_T^{\rm SK}/M_T^{\rm Kamioka})\sim\mathcal{O}(10^3)$ 
consistent with Fig.~\ref{fig:int_BDM_event}.
Super-K here only starts to accumulate events for $m_\chi\gtrsim 1.1$~keV whose $t_{\rm van} > 9~{\rm yr}$, and eventually dominates the contribution to $N_\chi$ more than that from Kamiokande for larger $m_\chi$.
For comparison, we also plot the combined background numbers $N_b$ from both detectors. 

\begin{figure}
\begin{centering}
\includegraphics[width=1\columnwidth]{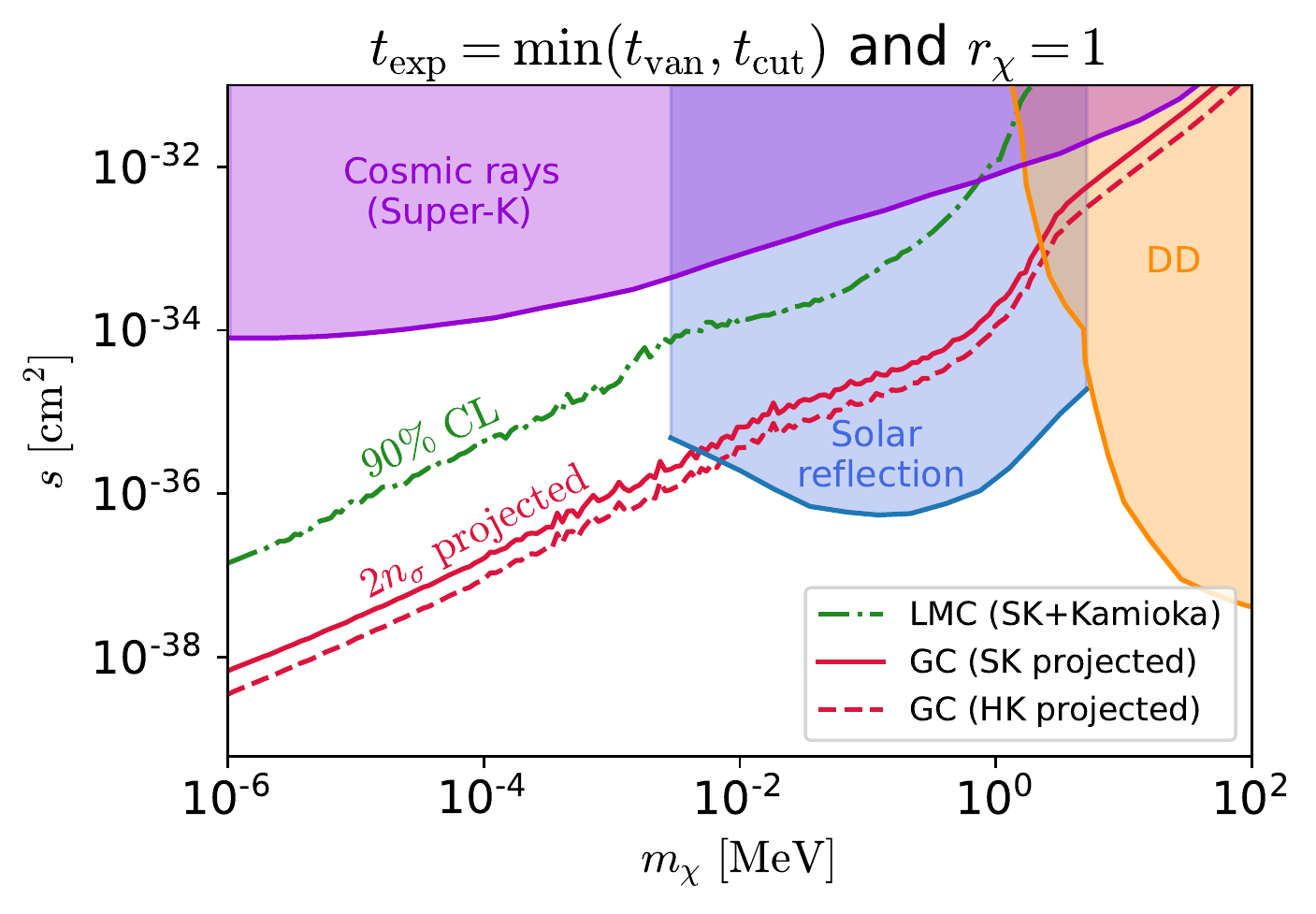}
\par\end{centering}
\caption{\label{fig:sensitivity} 
Sensitivity of BDM searches in $(m_\chi – s )$ plane. $s = \sqrt{\sigma_{\chi\nu} \sigma_{\chi e}}$ for this work on
the experimental limits at $n_\sigma=1.64$ (90\% CL) for LMC (green) and
projected $s$ for the GC at $n_\sigma = 2$ (red).
Current leading limits from cosmic-ray BDM~\cite{Cappiello:2019qsw},
solar reflection~\cite{An:2017ojc}, and DD~\cite{XENON:2019zpr,XENON:2019gfn,SENSEI:2019ibb,SuperCDMS:2018mne} at $s=\sigma_{\chi e}$ are superimposed.}
\end{figure}

{\it Sensitivity and constraint.---}We use $N_\chi$ and $N_b$ derived above to estimate the constraint and sensitivity on light DM, taking for simplicity 
\begin{equation}
    n_\sigma=\frac{N_\chi}{\sqrt{N_\chi+N_b}}.\label{eq:sensitivity}
\end{equation}
The dependencies of sensitivity ($s$) versus $m_\chi$ are
displayed in Fig.~\ref{fig:sensitivity}, where
$s = \sqrt{\sigma_{\chi\nu} \sigma_{\chi e}}$ for the experimental limit
at $n_\sigma=1.64$ [90\% confidence level (CL)] for SN1987a in LMC, and the
projected sensitivity at $n_\sigma=2.0$ for a SN in the GC.
In order to compare $s$ with existing constraints based exclusively on $\sigma_{\chi e}$ \cite{Cappiello:2019qsw,An:2017ojc,XENON:2019zpr,XENON:2019gfn,SENSEI:2019ibb,SuperCDMS:2018mne}, a model-dependent choice relating $\sigma_{\chi \nu}$ and $\sigma_{\chi e}$ has to be made\footnote{Our bounds are considerably better than those reported in Ref.~\cite{Jho:2021rmn,Das:2021lcr}, which are not shown in Fig.~\ref{fig:sensitivity}.  Also noted is that such a comparison implicitly assumes a cross section that is independent of the center-of-mass energy, which was adopted similarly in pioneer works for cosmic-ray upscattered dark matter scenario~\cite{Bringmann:2018cvk,Ema:2018bih,Cappiello:2019qsw}.}. Under a generic description of $r_{\chi} = \sigma_{\chi \nu}/\sigma_{\chi e}$, the specific case of $r_\chi=1$ was selected as illustration, with which the resulting bounds are superimposed in Fig.~\ref{fig:sensitivity}.
The Super-K constraints are derived from the average background rates \cite{Super-Kamiokande:2016yck}
and statistical uncertainties. Time stability can be inferred from the absence of
anomalous time variations in the solar $\nu$ annual modulation analysis \cite{Super-Kamiokande:2003snd,Super-Kamiokande:2008ecj}.
Limits derived with BDM from SN1987a in LMC leads to orders of magnitude improvement over existing bounds for $m_\chi < 2\,{\rm keV}$ over a large range of $r_\chi$.
For instance, more stringent limits are derived at $m_\chi\sim 10^{-6}$ MeV for $r_\chi>10^{-6}$.
Moreover, a future SN in the GC can improve the sensitivity by a factor of $\sim 30$ with Super-K, since $s^{\rm GC}_\chi/s^{\rm LMC}_\chi\sim \sqrt{N^{\rm LMC}_\chi/N^{\rm GC}_\chi}\sim \mathcal{O}(0.03$), allowing one to probe $s\lesssim 10^{-36}$~cm$^2$ for $m_\chi\lesssim 10$~keV. 
For $m_\chi\lesssim 100\,{\rm keV}$, the sensitivity curves follow $s \propto m_\chi^{1/2}$ simply due to $N_\chi\propto m_\chi^{-1}$ (see Fig.~\ref{fig:int_BDM_event}).

On the other hand, the sensitivities for $m_\chi\gtrsim 100\,{\rm keV}$ weaken considerably 
due to the reduced BDM that can arrive at the detector within 35 years.
Finally, we include an additional projection with Hyper-K for the GC case (red-dashed curve).
The analysis is similar to Super-K, with fiducial mass and background rate scaled up by a factor of 10, which then leads to another improvement of $\sim$ 2--3 over the Super-K result.

{\it Summary and prospects.---}We have examined the scenario of halo DM being boosted
by prompt SN$\nu$, and extracted a wealth of information from its TOF measurements.
The BDM events on Earth are characterized by unique timing distributions, which
vanish beyond $m_\chi$-dependent end points and are independent of the interaction
cross sections, while their peak positions provide information on the SN locations
and $m_\chi$.

A new constraint was derived on $s=\sqrt{\sigma_{\chi\nu}\sigma_{\chi e}}$ using Kamiokande and Super-K data on the SN1987a in LMC.
Our results probe and exclude new parameter space over a large range of $r_\chi$ and in
particular improve over the existing cosmic-ray BDM bounds for $m_\chi<100$~keV by 1--3
orders of magnitude at $\sigma_{\chi e}\sim\sigma_{\chi \nu}$. A future SN in the GC can provide improved
sensitivity by another factor of 30--100 with Super-K or Hyper-K. The improvement
over other probes~\cite{Cappiello:2019qsw,An:2017ojc,XENON:2019zpr,XENON:2019gfn,SENSEI:2019ibb,SuperCDMS:2018mne,Jho:2021rmn,Das:2021lcr} in the sub-MeV mass range originates from
the transient BDM flux arriving in a short duration that can be calibrated by the
detection of SN$\nu$, thereby minimizing the background counts. The constraint and
sensitivity of this work were derived by a conservative analysis which stands on the
BDM rates not being larger than those of background. A detailed analysis that
optimally exploits the $m_\chi$-dependent TOF temporal profile or combines multiple
detectors is beyond the scope of this work but will further enhance the sensitivities.
Furthermore, most BDM arriving on Earth are within a small solid angle relative to
the SN direction for $m_\chi \lesssim \mathcal{O}({\rm MeV})$~(see Fig.~\ref{fig:f}). Coupled with the good pointing capability for galactic SN~\cite{Beacom:1998fj,Tomas:2003xn,Super-Kamiokande:2016kji,Linzer:2019swe}, the angular information can be exploited to greatly reduce the background.

Other effects such as the distortion of SN$\nu$ spectra, the recently proposed SN$\nu$ echo~\cite{Carpio:2022sml}, and the impact of $\chi-\nu$ interaction on SN$\nu$ emission have been neglected here. Estimations suggest that SN$\nu$ spectra be minimally affected for the
parameter space examined.
These effects may be combined with the TOF profiles of
SN$\nu$BDM to provide severe constraints on specific phenomenological models relating
$\sigma_{\chi e}$ and $\sigma_{\chi\nu}$. With all the rich information, the next galactic SN will offer new insights to the nature of DM. Furthermore, TOF analysis following SN or other
transient astrophysical events can be applied in a similar vein to studies of other
exotic physics interactions.
A broad range of interesting scenario will be explored in our future research.

{\it Code availability.---}We provide a Python package \texttt{snorer} \cite{snorer2024}, which can fully reproduce the results in this {\it Letter}. The package is available on both GitHub and PyPI. It offers numerous new features, such as including DM spikes, user-specified SN locations in arbitrary distant galaxies, and an implementation of any particle physics model. See its official page for further details.

\begin{acknowledgments}
We thank Gang Guo, Kin-Wang Ng, Gianluca Petrillo, Yun-Tse Tsai for useful discussions.
Y.-H.~L.~acknowledges the Postdoctoral Scholar Program of the Academia Sinica, Taiwan, during which the major part of this work was done, as well as supports from National Science and Technology Council, the Ministry of Education under Project No.~NTU-112L104022, and the National Center for Theoretical Sciences of Taiwan.
M.-R.~W.~acknowledges supports from the National Science and Technology Council, Taiwan under Grant No.~110-2112-M-001-050 and 111-2628-M-001-003-MY4, the Academia Sinica under Project No.~AS-CDA-109-M11, and Physics Division, National Center for Theoretical Sciences of Taiwan.
H.~T.-K.~W. acknowledges supports from the National Science and Technology Council, Taiwan under Grant No.~106-2923-M-001-006-MY5. 
We also acknowledge the computing resources provided by the Academia Sinica Grid-computing Center.
\end{acknowledgments}


%

\end{document}